\documentclass[aps,prl,twocolumn,superscriptaddress]{revtex4-2}
\usepackage{amsmath}
\usepackage{graphicx}

\begin{document}


\title{Direct observation of magnon BEC in an out-of-plane magnetized yttrium iron garnet film}




\author{G. A. Knyazev}
\affiliation{ 
	M-granat, Russian Quantum Center, Skolkovo, Bolshoy Bulvar, 30, 121205 Moscow, Russia
}%
\affiliation{ 
	Photonic and Quantum technologies school, Lomonosov Moscow State University, Leninskie gori, 119991 Moscow, Russia
}%

\author{A. N. Kuzmichev}
\affiliation{ 
	M-granat, Russian Quantum Center, Skolkovo, Bolshoy Bulvar, 30, 121205 Moscow, Russia
}%

\author{P. E. Petrov}
\affiliation{ 
	M-granat, Russian Quantum Center, Skolkovo, Bolshoy Bulvar, 30, 121205 Moscow, Russia
}%
\affiliation{ 
	Photonic and Quantum technologies school, Lomonosov Moscow State University, Leninskie gori, 119991 Moscow, Russia
}%

\author{I. V. Savochkin}
\affiliation{ 
	M-granat, Russian Quantum Center, Skolkovo, Bolshoy Bulvar, 30, 121205 Moscow, Russia
}%

\author{P. M. Vetoshko}
\affiliation{ 
	M-granat, Russian Quantum Center, Skolkovo, Bolshoy Bulvar, 30, 121205 Moscow, Russia
}%
\affiliation{ 
	Kotelnikov Institute of Radioengineering and Electronics, Mokhovaya 11-7, 125009 Moscow, Russia
}%

\author{V. I. Belotelov}
\affiliation{ 
	M-granat, Russian Quantum Center, Skolkovo, Bolshoy Bulvar, 30, 121205 Moscow, Russia
}%
\affiliation{ 
	Photonic and Quantum technologies school, Lomonosov Moscow State University, Leninskie gori, 119991 Moscow, Russia
}%

\author{Yu. M. Bunkov}
\email{y.bunkov@rqc.ru.}
\affiliation{ 
	M-granat, Russian Quantum Center, Skolkovo, Bolshoy Bulvar, 30, 121205 Moscow, Russia
}%
\date{\today}

\begin{abstract}
Bose-Einstain condensation occurs at an appropriate density of bosonic particles, depending on their mass and temperature. We were able to experimentally observe the transition from the spin wave regime to the magnon Bose-Einstein condensed state (mBEC) with increasing magnon density by a microwave pumping. We used optical methods to register the spatial distribution of the magnon density and phase. For the first time, a coherent state of stationary magnons was demonstrated far from the region of their excitation.

\end{abstract}


\maketitle

Magnetism is, in principle, a quantum phenomenon, which is usually described in the semiclassical approximation. However, there are a number of phenomena to which the semiclassical consideration is not applicable. And first of all, this is Bose - Einstein condensation of magnons --- elementary excitations of the ground magnets state. It follows from quantum statistics that magnons should form a coherent quantum state (Bose-Einstein condensed state, mBEC) at a concentration above the critical value $N_c$. Magnons density is determined by the temperature and under stationary conditions it is always below  $N_c$. However, the density of magnons can be significantly increased by exciting them by radio-frequency (RF) photons. This process corresponds to the magnetic resonance --- deflection  and precession of magnetization in the quasi classical model of magnetism. In this paper, we study the properties of magnons with $\vec{k}=0$ in an out-of-plane magnetized YIG film at room temperature. Under these conditions, the homogeneous precession corresponds to the energy minimum and does not decay into spin waves \cite{Stamp}. The critical magnon concentration  $N_c$ for this case was calculated in \cite{BunkovSafonov} and corresponds to a deviation of the precessing magnetization by about 3$^\circ$.

In this article, we restrict ourselves to considering only mBEC for stationary magnons excited by resonant microwave pumping, whose properties are similar to those of an atomic Bose condensate. The coherent state of traveling magnons observed in in-plane magnetized YIG films has completely different properties and is not considered in this article.

Bose condensation of stationary magnons was previously discovered in antiferromagnetic superfluid $^3$He-B \cite{HPD}. It leads to the formation of a long-lived induction signal, which decayed orders of magnitude slower than it should be due to the inhomogeneity of the magnetic field. Spontaneous recovery of coherence after the decay of homogeneous precession \cite{spontan}, as well as a thousandfold narrowing of the resonance line \cite{nerrow}, clearly indicated the formation of mBEC state. The discovered phenomenon of magnon supercurrent \cite{super} is also a consequence of Bose condensation.

Despite the fact that antiferromagnetic $^3$He is a superfluid liquid, its superfluidity does not play any role in the formation of mBEC and magnon supercurrent. Superfluid properties are not included in any of the mBEC parameters. Thus, mBEC similar to that obtained in $^3$He can also be observed in solid magnets. In particular, the properties of magnons in $^3$He have many analogies with magnons in a YIG film \cite{mBEC4}. Magnons in an out-of-plane magnetized YIG film are characterized by repulsion, as in $^3$He-B, which leads to an upward frequency shift when the magnetization deviates. Therefore, mBEC can be similar to that in $^3$He-B.

The main advantage of antiferromagnetic superfluid $^3$He is the extremely long lifetime of magnons. The Gilbert damping constant is of the order of 10 $^{-8}$, which makes it possible to observe Bose magnon condensation after turning off the RF excitation. For YIG  the Gilbert constant is by 3 orders of magnitude larger, which makes it problematic to observe the formation of mBEC after turning off the RF excitation. 

In this article, we used another method for studying magnons, which also categorically confirms the formation of mBEC in a YIG film. Using an optical setup, we observed the spatial distribution of the magnon state outside the region of their excitation. These experiments showed the formation of mBEC at magnon concentrations above the critical value in the region where there is no RF excitation.

\section{Experimental setup}

\begin{figure}[h!]
	\centering\includegraphics[width=7cm]{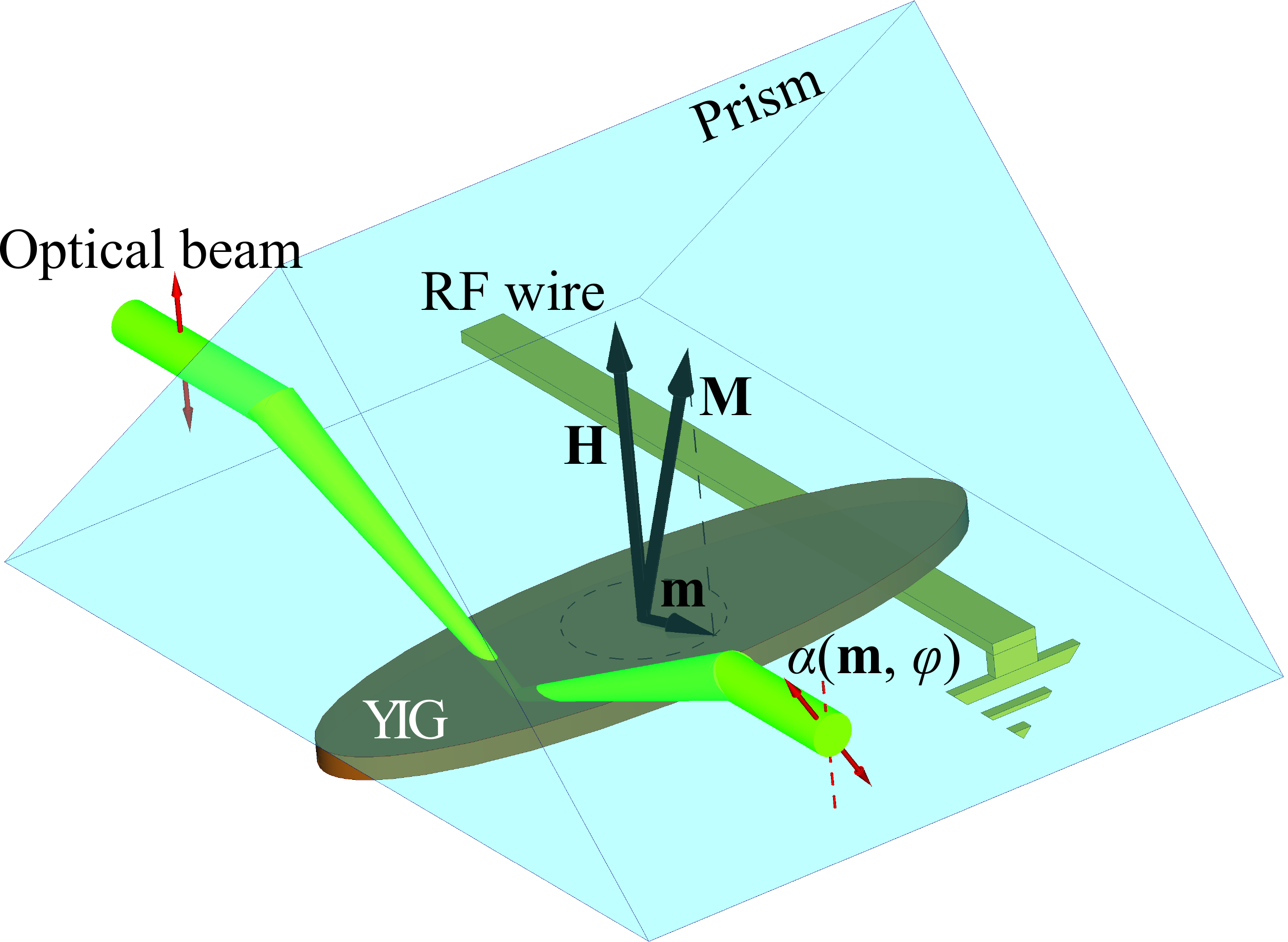}
	\caption{The scheme of  the experiment geometry. A beam of linearly polarized light illuminates the sample through a prism. The reflected beam is directed on the detectors. The polarization of the optical beam changes due to the Faraday rotation on the  magnetization component {\boldmath M} along the light beam. The received signal contains information about  the angle of magnetization  deflection $\Theta$ and phase of precession $\phi$.}
	\label{YigVec}
\end{figure}

The experiments were carried out on an elliptical YIG film 6 $\mu$m thick and 4.5 × 1.5 mm in size. The geometry of the experiment is schematically shown in  Fig. \ref{YigVec}. The sample was grown by epitaxy on a gallium gadolinium garnet substrate.
The elliptical shape of the sample was chosen to reduce the possibility of the secondary resonance modes formation. Magnetic resonance was excited by a narrow strip line 0.2 mm wide, oriented perpendicular to the main axis of the sample.
It was located at a distance of 1 mm from one of the sides of the sample. 

The states of magnons along the main axis of the sample were studied under local excitation of magnons by the radio-frequency pumping. A beam of linearly polarized light was
sent to the sample through a prism. The position of the laser beam was scanned along the sample. After the reflection of the beam, its optical polarization changes due to the Faraday rotation when interacting with the  component of the magnetization {\boldmath M} along light beam. Therefore, the Faraday angle is sensitive to the magnetization dynamics, in particular to its deflection angle $\Theta$ and phase $\phi$. 
The reflected beam of light was directed to a balance photodetector through a Wollaston prism.
Therefore, the signals from detectors carried information about the amplitude and phase of the precessing magnetization. To translate studies to lower frequencies, we used light modulation at a frequency shifted from the resonant frequency by about 12 kHz. This made it possible to record the parameters of the reflected signal using low-frequency detectors. After detecting the signals, we obtained the amplitude and phase of the magnon precession at a given point of the sample and for a given magnetic field. By scanning the magnetic field and the illumination point of the sample, we obtained the amplitude and phase distribution of magnons in the sample depending on the position and field. 
A detailed description of our optical setup was presented in
\cite{Petrov2022} where the state of magnons in the region of excitation  was investigated.

\section{Experimental results}

\begin{figure*}[]
	\includegraphics[width=16cm]{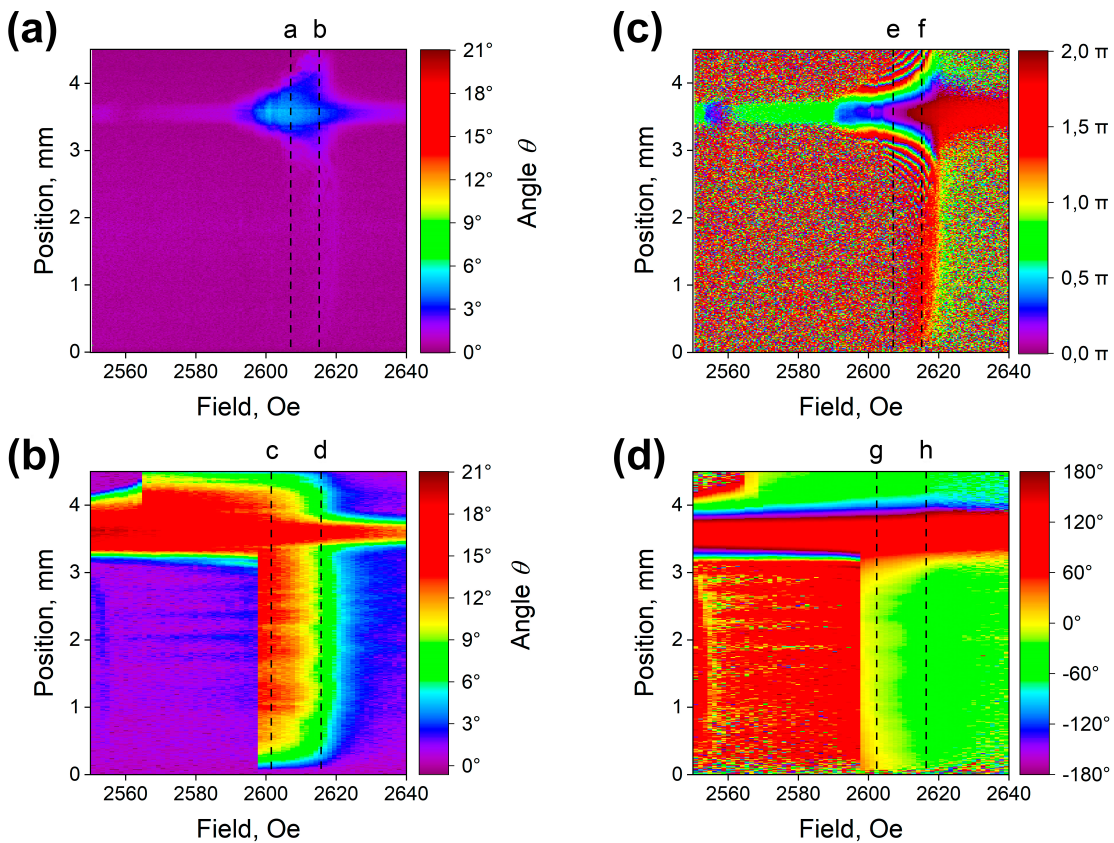}
	\caption{(a)--(b) The spatial distribution of magnon density in units of magnetization deflection angle as a function of sweeping down magnetic field at 0.05 mW  (a), and  at 6 mW (b) of pumping energy. (c)--(d) The spatial distribution of the magnon phase  as function of sweeping down magnetic field at 0.05 mW  (c) and at 6 mW (d) pumping energy.}
	\label{3DAP}
\end{figure*}

In this article, we demonstrate  the experimental observation of magnons state outside the excitation region. Experimental records of spatial destribution of  amplitude and phase of the magnetization precession as a function of the magnetic field at different excitation energies are shown in the figures \ref{3DAP}.
 

In Fig. \ref{3DAP} (a,b) the amplitude of signal in units of magnetization deviation $\Theta$ at a low excitation level of 0.05 mW (a) and at a high excitation level of 6 mW (b) is shown. The dependences of the signal amplitude on the magnetic field shift and the position of testing laser beam are shown. The deviation angle $\Theta$ was calibrated from the field shift from the resonance. The excitation strip line was located between the positions of 3.5 and 3.7 mm. At the low excitation, the resonance  is clearly visible with a maximum in the region of the strip line and a resonance field of about 2621 Oe and slightly below.

This shift from resonanse takes place due to effect of ``Foldover" rsonanse, when the resonanse field shifts due to decrease of demagnetization field at magnetization deflection.  See \cite{Petrov2022,Bunkov2021} for details. 
The spatial distribution of magnons changes dramatically at a higher level of excitation. Magnons filled the entire sample in the range of fields from 2621 to 2598 Oe. At a lower field, magnons do not fill the part of the sample below the strip line, but continue to fill the upper part of the sample up to 2562 Oe. This field shift is determined by magnon relaxation, which is proportional to the square of the deflection angle and the distance from the excitation region to the edge of the sample. The filling with magnons disappears as soon as their flow from the excitation region ceases to compensate for the losses.

Of great interest is the spatial distribution of the precession phase in these experiments, shown in Fig. \ref{3DAP} (c,d). We see that at low excitation power magnons propagate from the excitation region in the form of spin waves. The length of these waves changes with the shift of the magnetic field. Naturally, the magnetization precession frequency should coincide with the RF pump frequency. In the absence of an additional radio frequency field, the frequency shift is provided by the gradient energy of spin waves. Therefore, as the magnetic field decreases, the length of the spin waves also decreases. This effect is clearly seen in computer micromagnetic simulation of experimental conditions in the framework of the semiclassical Landau-Lifshitz theory.

This state changes completely with increasing RF pump energy, as shown in Fig. \ref{3DAP} (d). We see that a spatially homogeneous state with a coherent phase distribution arises outside the pump region. In this case, it should be emphasized that  coherence is a property of magnons, since it is not supported by an external RF field. This state directly demonstrates magnon Bose condensation at a high concentration of magnons. Indeed, Fig. \ref{3DAP}(d) clearly shows that magnons fill the entire sample at a magnon concentration corresponding to the magnetization deviation angle of more than 3$^\circ$ , as predicted in the work \cite{BunkovSafonov}. 

\begin{figure}[h!]
	\includegraphics[width=9cm]{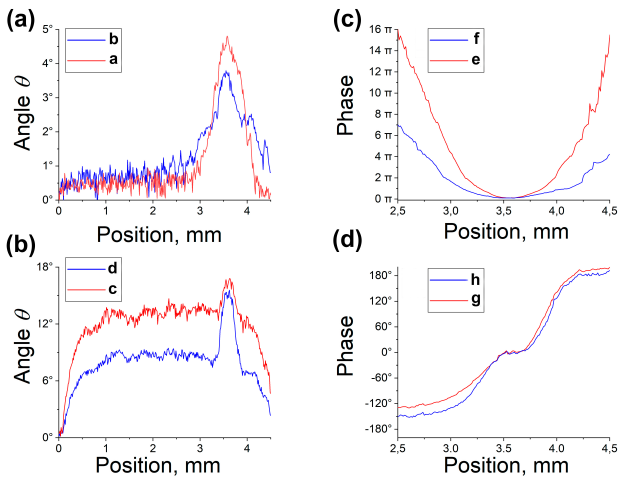}
	\caption{(a)--(b) The spatial distribution of the magnon density in units of the magnetization deflection angle in the fields corresponding to the dashed lines in Fig.\ref{3DAP} (a,b) at a pump energy of 0.05 mW (a) and  6 mW (b). (c)--(d) The spatial distribution of the magnon phase  at fields  corresponding to the dashed lines in Fig.\ref{3DAP} (c,d) at a pump energy of 0.05 mW (c) and  6 mW (d).}
	\label{Slices2DAP}
\end{figure}

Let's consider the spatial distribution of magnon amplitude and phase at a fixed field in more detail. In Fig. \ref{Slices2DAP} (a) the spatial distribution of the signal amplitude is shown for an external magnetic field of 2607 Oe and 2615 Oe, marked by a dashed lines in Fig. \ref{3DAP} (a), and  at an RF pump power of 0.05 mW. The critical magnon concentration is reached only in the RF pump region, while outside this region the magnon concentration is much lower and a magnon gas can be described in the semiclassical approximation. The very different destribution is shown in Fig. \ref{Slices2DAP} (b) at an excitation power of 6 mW. In this case, the magnons filled the entire sample in approximately the same concentration, excluding the edge regions.

The spatial distribution of the magnon precession phase is shown in Fig. \ref{Slices2DAP} (c,d). With a small excitation of 0.05 mW, a phase rotation is observed, which depends on the distance from the excitation region, which reaches 7$\pi$ per mm in a field of 2615 Oe and 17$\pi$ per mm  in a field of 2607 Oe. We calculated the spatial destribution of phase precessing for the experimental conditions by micromagnetic modeling using the MuMax3 \cite{Mumax} program and obtained excellent agreement with the experimental results.

The phase distribution changes drastically at 6 mW excitation as shown in Fig. \ref{Slices2DAP} (d). 
In this case, a sharp turn of the precession phase near the excitation region corresponds to the flow of magnons from the excitation region. For small magnon relaxation, this turn corresponds to about 180$^\circ$. It decreases with increasing relaxation. This experimental result requires further theoretical study. It is important to note that micromagnetic modeling by MuMax3 program leads to the formation of waves at any excitation amplitude. We can conclude that the formation of a spatially homogeneous coherent precession at high excitation is a consequence of magnons Bose  condensation, which, naturally, lies beyond the scope of the semiclassical theory. The MuMax3 program is based on the classical theory of Landau-Lifshitz magnetization precession and, of course, cannot simulate Bose condensation, which has a quantum nature.
 

\section{Conclusion}

Magnon Bose condensation and the associated magnon supercurrent in antiferromagnetic superfluid $^{3}$He are well known quantum phenomena that have received worldwide recognition \cite{Bunkov2013} and are awarded by F. London Memory Prize for this discovery. The formation of a Bose condensate of stationary magnons in solid-state magnets has caused much controversy.
Several experimental results obtained earlier in the out of plane magnetized YIG film were considered as indirect evidence of magnons Bose  condensation \cite{Bunkov2021,Vetoshko2020}. In this article, magnon Bose  condensation is experimentally demonstrated by direct optical observation of the coherent precession of magnetization far beyond the excitation region. This result contradicts the semiclassical Landau-Lifshitz theory, which directly indicates the quantum nature of this effect. It opens up new perspectives for research in quantum physics, as well as for some modern technological applications of magnonics, quantum communications, and quantum computing.

This work  was supported by Rosatom in the framework of the Roadmap for Quantum computing (Contract No. 868-1.3-15/15-2021 dated October 5).


\begin{thebibliography}{55}
	
\bibitem{Stamp} Tupitsyn, I. S., Stamp, P. C. E., \& Burin A.L. ``Stability of Bose-Einstein Condensates of Hot Magnons in Yttrium Iron Garnet Films". {\it Phys. Rev. Lett.} {\bf 100,} 257202 (2008)
	
	\bibitem{BunkovSafonov} Yu. M. Bunkov and V. L. Safonov,
``Magnon condensation and spin superfluidity".
{\it J. Mag. and Mag. Mat.} {\bf 452,} 30 (2018).
	
	\bibitem{HPD} A. S. Borovik-Romanov, Yu. M. Bunkov, V. V. Dmitriev and Yu. M. Mukharskii,
``Long-lived induction signal in superfluid $^{3}$He-B".
{\it JETP Lett.} {\bf 40,} 1033 (1984).

\bibitem{spontan} Yu.M. Bunkov and G. E. Volovik
“Magnon Bose Einstein Condensation  and Spin Superfluidity
{\it J. Phys.: Condens. Matter} {\bf 22} 164210  (2010).


\bibitem{nerrow} Yu. M. Bunkov,  “Helium-3; Cosmological and atomic physics experiments”  {\it Phil. Trans. R. Soc. A}, {\bf 366}, 2821 (2008). 

\bibitem{super} A. S. Borovik-Romanov, Yu. M. Bunkov, V. V. Dmitriev, Yu. M. Mukharskiy,
D. A. Sergatskov, ''Investigation of Spin Supercurrent in 3He-B",
{\it Phys. Rev. Lett.}, {\bf 62}, 1631 (1989).

		
\bibitem{mBEC4} G.E. Volovik, "Twenty Years of Magnon Bose Condensation and Spin
Current Superfluidity in 3He-B", {\it J. of  Low Temp. Phys.}, {\bf 153}, 135 (2008).



 \bibitem{Petrov2022} P. E. Petrov, P. O. Kapralov, G. A. Knyazev, A. N. Kuzmichev, P. M. Vetoshko, Yu. M. Bunkov, V. I. Belotelov, "Magneto-optical imaging of coherent spin dynamics in ferrites" Optics Express 30, 1737 (2022).
 
 \bibitem{Bunkov2021} Yu. M. Bunkov, A. N. Kuzmichev, T. R. Safin, P. M. Vetoshko, V. I. Belotelov, and M. S. Tagirov, ``Quantum paradigm of the foldover magnetic resonance’’ {\it Scientific Reports}, {\bf 11}, 7673 (2021).
 
\bibitem{Mumax} A. Vansteenkiste et al., ''The design and verification of MuMax3"
{\it AIP Advances}, {\it 4}, 107133 (2014).
 
 
 \bibitem{Bunkov2013} Yu. M. Bunkov and G. E. Volovik,   ``Spin Superfluidity and Magnon BEC'' in {\it Novel Superfluids Ch.4,} (eds. Bennemann, K. H. \& Ketterson, J. B. Oxford Univ. Press, Oxford, (2013).
 
 \bibitem{Vetoshko2020} P. M. Vetoshko, G. A. Knyazev A. N. Kuzmichev, A. A. Holin, V. I. Belotelov, and Yu. M. Bunkov ``Bose condensation and spin superfluidity of magnons in a perpendicularly magnetized film of yttrium iron garnet'',  {\it JETP Lett.}, {\bf 112},  299 (2020).
 
 
\end{thebibliography}
\end{document}